\documentclass{ifacconf}

\usepackage{graphicx}      % include this line if your document contains figures
\usepackage{natbib}        % required for bibliography
\usepackage{siunitx}
\usepackage{amssymb}
\usepackage{mathtools}
\DeclarePairedDelimiter\abs{\lvert}{\rvert}
\DeclarePairedDelimiter\norm{\lVert}{\rVert}

\newtheorem{assumption}{Assumption}

\usepackage{enumitem}
\newlist{steps}{enumerate}{1}
\setlist[steps, 1]{label = \textbf{Step \arabic*:}}
\begin{document}
\begin{frontmatter}

\title{A Model-Free Loop-Shaping Method based on Iterative Learning Control \thanksref{footnoteinfo}} 

\thanks[footnoteinfo]{This work was financially supported from the Young Scholar Fellowship Program by Ministry of Science and Technology (MOST) in Taiwan, under Grant MOST 108-2636-E-002-007.}

\author[First]{Li-Wei Shih} 
\author[Second]{Cheng-Wei Chen} 

\address[First]{Department of Electrical Engineering, National Taiwan University, Taipei, Taiwan (e-mail: b04502031@ntu.edu.tw).}
\address[Second]{Department of Electrical Engineering, National Taiwan University, Taipei, Taiwan (e-mail: cwchenee@ntu.edu.tw).}

%================================================================================
\begin{abstract}                % Abstract of not more than 250 words.

 Many techniques have been developed for the loop-shaping method in control design. While most loop-shaping methods apply a model of the open-loop controlled plant, the resulting performance depends on the accuracy of the dynamical model. This paper aims to develop a model-free loop-shaping technique. The core idea is to convert the model matching problem to a trajectory tracking problem. To achieve the desired loop gain, we need to determine the control input such that the system output tracks the impulse response of the loop gain function. In this paper, a model-free iterative learning control (ILC) algorithm is applied to solve this tracking problem. Once the ILC converges, the feedback controller that meets the desired loop gain can then be constructed. This method does not require the model of the controlled plant, hence it provides better performance of loop-shaping control design. The proposed method is validated through numerical simulation on a 3-rd order plant. 
\end{abstract}


\end{frontmatter}

%================================================================================
\section{Introduction}
\label{sec:1}
To achieve closed-loop tracking performance, noise rejection, and robustness, loop-shaping has been applied to modern controller design in many applications [\cite{prempain2005static}], [\cite{zhu2003robust}], [\cite{fujita1993loop}]. Several techniques have been developed, such as PID-controller [\cite{grassi1996pid}] and $\mathcal{H}_\infty$ synthesis loop-shaping method [\cite{mcfarlane1992loop}]. The well-known $\mathcal{H}_\infty$ synthesis loop-shaping is wildly used to design a controller by assigning several weighting functions indicating the requirement of the desired performance and noise suppression.

The abovementioned work needs a parametric model of the open-loop controlled plant to work on. That is, dynamical modeling or system identification should be properly performed before applying these model-based loop-shaping methods. The resulting performance relies on the quality of the provided model. To eliminate the procedure of model construction, several types of research have been working on model-free loop-shaping approaches, such as a data-driven PID controller [\cite{saeki2014data}]. This method is useful for most low-order plants while it is still challenging to compensate for higher-order dynamics. Another approach is utilizing Linear Matrix Inequality (LMI) to design fixed-order controller with transient response data [\cite{saeki2008data}], [\cite{formentin2012data}]. Formulated as a convex optimization problem, LMI ensures a globally optimal solution. However, the computation is time-consuming when the variable dimension is large, especially when it is applied to transient response data. %In addition, for controller candidates, it can be validates whether it violates the specification directly using input-output data [\cite{sung2018data}].

Iterative Learning Control (ILC), on the other hand, has been shown its excellent capability of tracking [\cite{bristow2006survey}]. By iteratively working on the same task repeatedly, updating the current input with respect to the output error in the previous run, the input converges and generates the desired output. The learning filter is a critical design parameter in the ILC algorithm, which exerting influence on the convergence property of ILC. Model-free ILC has been proposed, such as PD-type updating law [\cite{chen2002optimal}], reverse-time technique [\cite{ye2005zero}], and model-free inversion-based ILC (MIIC) [\cite{kim2012modeling}]. As to increase the convergence rate, [\cite{chen2017data}] proposed a convergence rate acceleration algorithm. The algorithm learns a feedforward filter and crossly updates the learning filter from the on-learning feedforward filter during the iterative process. This algorithm can be applied to construct the plant inversion FIR filter in only a few iterations. The constructed plant inversion FIR filter is able to work as a learning filter to other ILC applications.

The main contribution of this paper is the development of a model-free loop-shaping method based on ILC. The conventional loop-shaping problem is converted to a tracking problem. The impulse response of the target loop gain function is assigned as the system's desired reference, and then the ILC algorithm is applied to track the reference at the system's output. Once the iterative learning process converges, the output response of the open-loop plant is the same as the impulse response of the target loop gain. The learned input is then used to construct the closed-loop controller. Therefore, the constructed controller along with the controlled plant yields the desired loop gain. A parametric model is unneeded in this method, resulting in better performance of loop-shaping controller design.

The remainder of this paper is organized as follows: Section \ref{sec:2} gives the problem definition and theoretical background; Section \ref{sec:3} demonstrates the proposed method for loop-shaping; simulation validation is presented in Section \ref{sec:4}; the conclusion is given in Section \ref{sec:5}.
%================================================================================
\section{Problem Definition and Preliminaries}
\label{sec:2}

%--------------------------------------------------------------------------------
\subsection{Loop-shaping}
For a typical loop-shaping problem, considering a SISO linear time-invariant(LTI) plant $P(z)$, which is in Z-transform representation as

\begin{equation} \label{eq:plant_def}
    P(z) = \sum ^{N/2}_{k=-N/2}p(k)z^{-k}
\end{equation}

where $p(k)$ is the impulse response of the plant $P(z)$ and $N \rightarrow \infty$. In the meanwhile, several closed-loop system behavior specifications are also assigned such as tracking performance, noise rejection, and robustness.
\begin{assumption}
    The Z-transform representation plant $P(z)$ is a proper minimum phase transfer function.
\end{assumption}
The purpose of loop-shaping is to find a controller $C(z)$ such that the overall loop gain $L(z)$ is equal to the desired loop gain $L_d(z)$ that satisfies the requirements. Note that both the controller $C(z)$ and the loop gain $L(z)$ are in the Z-transform representation, and the loop gain $L(z)$ is the series connection of the controller and the plant, i.e.,

\begin{equation} \label{eq:L}
    L(z) = C(z)P(z)
\end{equation}

For a desired loop gain $L_d(z)$ that is selected to meet the system specifications, the quality of the controller $C(z)$ designed can be verified by the error between $L_d(z)$ and the loop gain $L(z)$ corresponding to $C(z)$. Therefore, the goal is to find a controller $C(z)$ that minimizes the loop gain error. In other words,

\begin{equation} \label{eq:min_freq}
    \min_{C(z)}{\norm{L_d(z)-C(z)P(z)}_\infty}
\end{equation}

where $\norm{\cdot}_\infty$ denotes the $\mathcal{H}_\infty$-norm. This optimization cost function can also be converted to the time domain with the $\mathcal{H}_\infty$-norm substituting to $l^2$-norm.

\begin{equation} \label{eq:min_time}
    \min_{c(k)}{\abs{L_d(k)-c(k)P(z)}_2}
\end{equation}

Note that $\abs{\cdot}_2$ denotes the $l^2$-norm, $L_d(k)$, and $c(k)$ represent the impulse response of the desired loop gain $L_d(z)$ and the controller $C(z)$, respectively. Thus, the original loop-shaping problem is translated into a tracking problem form.
%--------------------------------------------------------------------------------
\subsection{Iterative Learning Control (ILC)}

For a control input $u(k)$ and the corresponding output $y(k)$ passing through a plant $P(z)$, ILC iteratively learns to track a reference tracking $r(k)$ by updates the control input in the following form:

\begin{equation} \label{eq:ilc}
    \begin{split}
    u_0(k) &= F(z)r(k) \\
    u_{j+1}(k) &= u_{j}(k) + F(z)[r(k)-y_j(k)] \\
    &= u_{j}(k) + F(z)e_j(k)
    \end{split}
\end{equation}

where $F(z)$ is the learning filter, and $e(k):=r(k)-y(k)$ is the tracking error. Note that the lower index $j$ denotes the $j$-th iteration in the ILC process.

With some simplification, the error at $j$-th iteration can be described as 
\begin{equation} \label{eq:ilc_err}
    e_{j+1}(k) = [1-P(z)F(z)]^{j+1}r(k)
\end{equation}

From Eq. \ref{eq:ilc_err}, it is observable that the iterative process converges if the following criterion is satisfied [\cite{norrlof2002time}]:

\begin{equation} \label{eq:ilc_conv}
    \gamma := \norm{I-P(z)F(z)}_\infty < 1
\end{equation}

Note that $\gamma$ is the error convergence rate. Eq. \ref{eq:ilc_conv} shows that the closer the multiplication of the plant $P(z)$ and the learning filter $F(z)$ is to \num{1}, the smaller the error convergence rate $\gamma$ is, which means faster convergence rate. Ideally, under the circumstance that $F(z)=P^{-1}(z)$, $\gamma$ is equal to \num{0}, which means the iterative learning process converges immediately at the first iteration. With respect to this idea, a satisfactory way to choose the learning filter is by finding the plant inversion $P^{-1}(z)$ as $F(z)$.
%--------------------------------------------------------------------------------
\subsubsection{ILC-based Feedforward Filter (ILCFF)}

To obtain a proper learning filter $F(z)$, the method proposed in [\cite{chen2019iterative}] can be applied to construct a learning filter through plant inversion. This approach utilized ILC to learn an inversion-based FIR feedforward filtered without constructing the parametric plant model by tracking a filter impulse. The ILC algorithm is used to minimize the following equation:

\begin{equation} \label{eq:ilcff_min_time}
    \min_{f(k)}{\abs{r(k)-f(k)P(z)}_2}
\end{equation}

where $r(k)$ is a target reference output and $f(k)$ is the corresponding learned feedforward filter coefficients. The reference output $r(k)$ is typically a filtered impulse filtered by a zero-phase low pass filter, avoiding control saturation for high gain occurring at high frequency. The inversion-based FIR feedforward filter is then constructed by

\begin{equation} \label{eq:ilcff_f}
    F(z) = \sum^{N/2}_{k=-N/2}f(k)z^{-k}
\end{equation}

Notice that the constructed inversion-based feedforward filter may be non-causal despite the relative order of the plant $P(z)$. To clarify, if the plant has a delay of order $n$, then its inversion filter has an advance of order $n$, resulting in a non-causal inversion filter.
%--------------------------------------------------------------------------------
\subsubsection{Convergence rate acceleration for ILC}
In comparison to the slow convergence rate in the conventional ILC methods, which use PD-type or reverse-time technique as a learning filter, an approach significantly improving the convergence rate for ILC was proposed in [\cite{chen2017data}]. The algorithm proposed in this approach updates the learning filter $L(z)$ from the learned feedforward filter $F(z)$ every constant iteration. Since the $F(z)$ learns the plant inversion FIR filter, each the learning filter is updated, the product $P(z)F(z)$ is closer to 1, leading to a faster convergence rate $\gamma$ presents in Eq. \ref{eq:ilc_conv}.
%================================================================================
\section{ILC-based loop-shaping}
\label{sec:3}

For the loop-shaping problem of the model matching form described in Eq. \ref{eq:min_freq}, we propose a model-free method to design the controller $C(z)$ utilizing ILC. The problem is converted into a time domain tracking problem states in Eq. \ref{eq:min_time}. As the key concept, the loop gain $L(z)$ constructed in Eq. \ref{eq:L} is equal to the desired loop gain $L_d(z)$ as long as the impulse response $l(k)$ learns the reference output $L_d(k)$. In other words, if there is a control input $c(k)$ to the plant $P(z)$ such that its output $l(k)$ is same as $L_d(k)$,
then the control input $c(k)$ can be used as an FIR filter shaping the loop gain. The method proposed in this paper can be separate to following four steps:
\begin{steps}[leftmargin=1.4cm]
	\item Design the desired loop gain function $L_d(z)$ according to given specifications.
	\item Generate the learning filter using ILCFF with convergence rate acceleration.
	\item Apply ILC to learn the impulse response of $L_d(z)$.
	\item Convert the learned FIR filter into an IIR filter using balanced model reduction.
\end{steps}
%--------------------------------------------------------------------------------
\subsection{Algorithm}
%--------------------------------------------------------------------------------
\textbf{Step 1} (Design of the loop gain function $L_d(z)$)

\begin{assumption}
    \label{asm:rel_order}
    The relative order of the desired loop gain $L_d(z)$ is higher than the relative order of the plant $P(z)$.
\end{assumption}

Once the design specification is provided, the desired loop gain $L_d(z)$ can be designed to meet the requirements. However, the limitation for the relative order relationship between $L_d(z)$ and $P(z)$, described in \textit{Assumption \ref{asm:rel_order}}, is necessary. The desired controller of the loop-shaping problem is denoted as $C_d(z)$, which can be evaluated by $C_d(z)=L_d(z)/P(z)$. If $C_d(z)$ is improper, the learned FIR filter $c(k)$ is non-causal. Non-causal FIR filters can only be realized with an additional input delay, which is not allowed in the implementation of feedback control. Hence for causal control input, $C_d(z)$ should be limited to be proper, resulting in the limitation between $L_d(z)$ and $P(z)$. To measure the relative order of the plant $P(z)$, a delta impulse is triggered as an input to the plant and the output response is recorded. Then the relative order of $P(z)$ is evaluated from the delay between the impulse input and the moment the output starting to response.
%--------------------------------------------------------------------------------

\textbf{Step 2} (Design of the ILC learning filter)

\begin{assumption}
    \label{asm:learning_filter}
    The learning filter $F(z)$ is chosen properly such that the ILC convergence criteria stated in Eq. \ref{eq:ilc_conv} is satisfied. Thus, the ILC algorithm converges to \num{0}, i.e. for any arbitrary small $\epsilon \in \mathbb{R}$, there exists $n^* \in \mathbb{N}$, such that $\abs{e_n(k)}_2 < \epsilon$ for all $n > n^*$, $n \in \mathbb{N}$.
\end{assumption}

While all kinds of learning filters that promise the convergence are suitable for the ILC process, while the error convergence rate $\gamma$ defined in Eq. \ref{eq:ilc_conv} depends on the quality of the learning filter chosen, however. To achieve a faster convergence rate, the learning filter generated from ILCFF with convergence rate acceleration method is suitable for the proposed ILC-based method. Taking advantage of its exponential error convergence rate acceleration, the method is practical to be implemented on a plant in the real world such that the experimental tasks required are affordable. 

%--------------------------------------------------------------------------------
% \subsection{ILC process}

\textbf{Step 3} (ILC for loop shaping design)

\begin{figure}[tb]
\begin{center}
    \includegraphics[width=8.4cm]{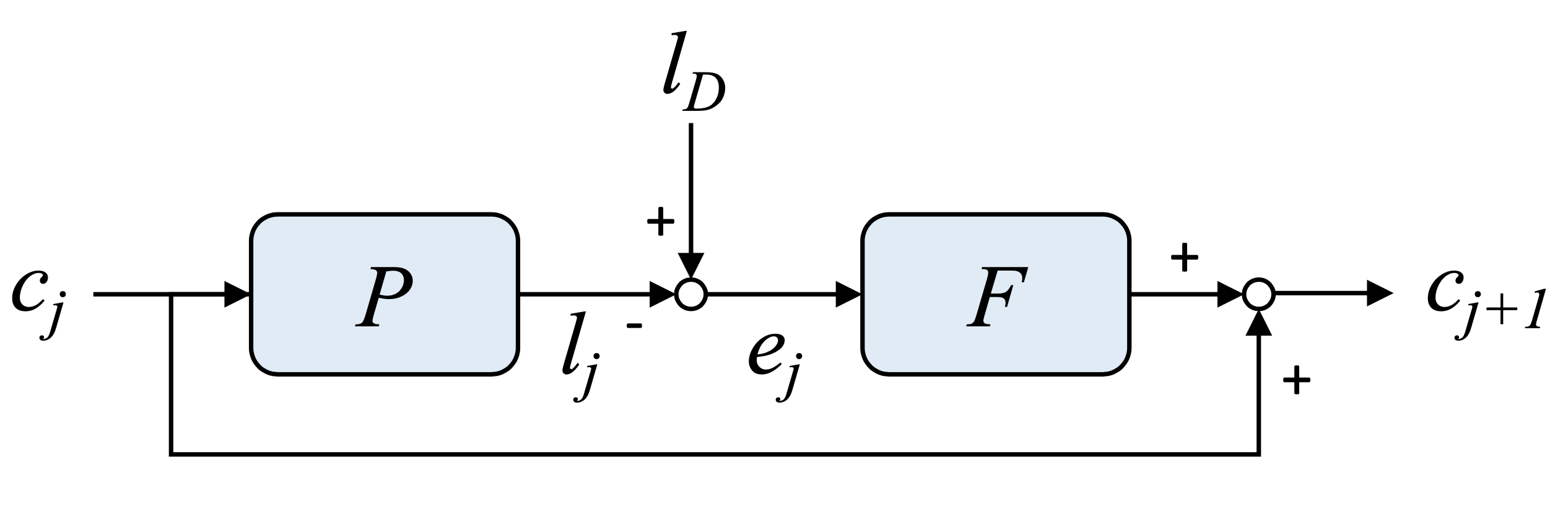}    % The printed column width is 8.4 cm.
    \caption{The block diagram for the propose method.}
    \label{fig:ILC}
    \end{center}
\end{figure}

With the constructed learning filter, the ILC algorithm is applied to the tracking problem in Eq.\ref{eq:min_time}. The block diagram is shown in Fig. \ref{fig:ILC} and the update law of iterative process is stated as

\begin{equation} \label{eq:ilc_ls}
    \begin{split}
    c_{j+1}(k) &= c_{j}(k) + F(z)[L_d(k)-l_j(k)] \\
    &= c_{j}(k) + F(z)e_j(k)
    \end{split}
\end{equation}

Since the selected learning filter satisfies \textit{Assumption \ref{asm:learning_filter}}, the ILC algorithm for the tracking problem converges. The corresponding learned FIR filter $C(z)$ is constructed as

\begin{equation} \label{eq:c}
    C(z) = \sum^{N/2}_{k=-N/2} \lim_{j \rightarrow \infty} c_j(k)z^{-k}
\end{equation}

where $N$ is the length of the learned input filter, and $N \rightarrow \infty$ ideally. This corresponds to an infinite long FIR filter, which is impractical for implementation. Hence the presence of truncation is necessary, the zero-closed term should be truncated at an appropriate threshold. An instinctive way is to consider the settling property of the impulse response of $P(z)$ and $L_d(z)$, where the acceptable error threshold is a design parameter depends on the implementation, selecting the longer one as the filter length $N$.

%--------------------------------------------------------------------------------
% \subsection{IIR filter conversion}
\textbf{Step 4} (Controller order reduction)

For real-time feedback control, there is a disadvantage of directly using the learned controller $C(z)$ constructed from the learned input $c(k)$. By selecting the threshold, the infinite long impulse response is truncated to a finite length $N$. Nevertheless, the length is still too long for real-time implementation which required considerable computation for filter convolution. This use of the FIR filter also results in that the closed-loop system needs at least $N$ samples to achieve steady state, causing poor behavior in transient response.

To achieve a better transient response, it is necessary to reduce the order of the learned controller $C(z)$. The controller then becomes an IIR filter denotes as $C_{\text{IIR}}(z)$, which can be obtained by model reduction. The error due to balanced model reduction, $\norm{C(z)-C_{\text{IIR}}(z)}_\infty$, is promised through its Hankel singular values [\cite{glover1984all}], it has been shown that

\begin{equation} \label{eq:err_bound}
    \norm{C(z)-C_{\text{IIR}}(z)}_\infty \le 2 \sum_{k\ge r+1}^N \sigma_k
\end{equation}

where $\sigma_k$, $k \in \mathbb{N}$, $0 < k \le N$, is the $k$-th Hankel singular value of the control input $C(z)$, and $r$ is the order of the reduced IIR filter. Specifies that $\sigma_1 \ge \sigma_2 \ge \hdots \ge \sigma_r \ge \hdots \ge \sigma_N$.

Once $C(z)$ is truncated by balanced model reduction, the IIR-type loop gain $L_{\text{IIR}}(z)$ is expressed as

\begin{equation} \label{eq:L_IIR}
    L_{\text{IIR}}(z) = C_{\text{IIR}}(z)P(z)
\end{equation}

Since he desired controller $C_d(z)$ can be estimated through directly division $C_d(z)=L_d(z)/P(z)$. The order of the IIR-type controller $C_{\text{IIR}}(z)$ can be evaluated with and no more than the sum of the orders of $L_d(z)$ and $P(z)$.

%--------------------------------------------------------------------------------

\subsection{Frequency Weight}
When the pole of the desired loop gain $L_d(z)$ getting closer to the unit circle, it takes longer for the impulse response of $L_d(z)$ to approaches \num{0}. This causes that the length $N$ of the impulse response should be chosen specific large in order to reduce the error due to truncation. It is even impossible when $L_d(z)$ contains an integrator, which needs an infinity long length and unable to be truncated.

To solve this problem, the desired loop gain $L_d(z)$ can be separated into two parts $H(z)$ and $L_d'(z)$. 

\begin{equation} \label{eq:freq_weight}
    L_d(z) = H(z)L_d'(z)
\end{equation}

where $H(z)$ is a frequency weighting function such as integrator or the other slow poles that cannot be truncated under the selected filter length $N$. $L_d'(z)$ contains all the remaining poles and zero. Therefore, the proposed ILC-based loop-shaping method can be applied to track the impulse response of $L_d'(z)$, denoted by $L_d'(k)$. The controller $c(k)$ can be reconstructed with the learned controller input $c'(k)$ and $H(z)$.

\begin{equation} \label{eq:c_h}
    c(k) = H(z) c'(k)
\end{equation}

Note that it is necessary to ensure $L_d'(z)$ is proper, otherwise its impulse response is unstable instead of converging to \num{0}.
While $L_d'(z)$ is improper due to the separation of $H(z)$ and $L_d'(z)$, Eq. \ref{eq:freq_weight} should be modified to create additional pole-zero pairs. To clarify, $L_d'(z)$ is multiplied by additional poles and hence becomes strictly proper. $H(z)$ is also multiplied by the additional zeros.
\begin{equation} \label{eq:integrator_alt}
    L_d(z) = (H(z) z^n) (L_d'(z) \frac{1}{z^n}) 
\end{equation}
where $n$ is the number of additional pole-zero pairs. The placement of the additional pole-zero pairs is located infinitely far for simplicity, hence the additional poles are promised not to make the settling property of tracking output slower. In addition, the frequency weight with an additional zero is stable since it is still a proper filter.

%================================================================================
\section{Simulation Validation}
\label{sec:4}
%--------------------------------------------------------------------------------
\subsection{Setup of the Simulation}

\begin{figure}[b]
    \begin{center}
    \includegraphics[width=8.4cm]{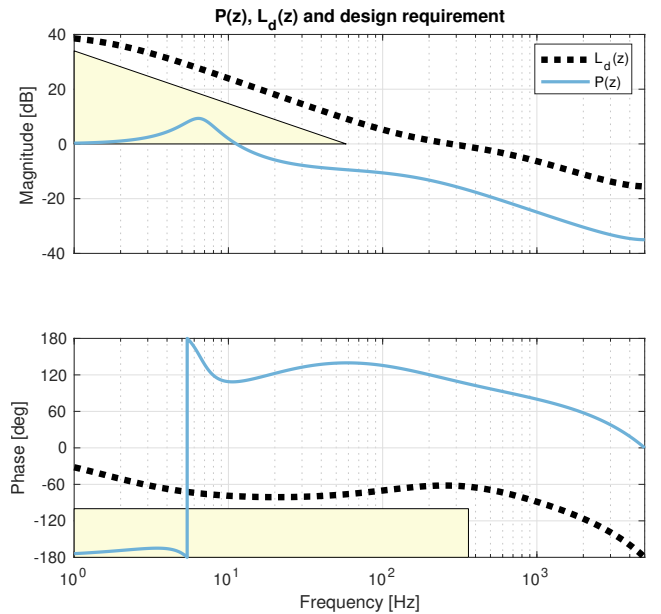}    % The printed column width is 8.4 cm.
    \caption{The Frequency response of the plant $P(z)$ and the desired loop gain $L_d(z)$. Forbidden regions for specifications are shown in the yellow area.} 
    \label{fig:P}
    \end{center}
\end{figure}

The performance of the proposed method is validated by numerical simulation. Considering a SISO 3-rd order plant P(z) defined as

\begin{equation} \label{eq:plant}
    P(z) = \frac{-0.1(z-0.995)(z-0.99)}{(z-0.4)(z^2-1.998z+0.998)}
\end{equation}

The frequency response of P(z) is presented in Fig. \ref{fig:P}. The plant has a bandwidth of \SI{14.2}{\Hz}. In addition, there is a peak response at \SI{6.51}{\Hz}. The controller sampling rate is assumed to be \SI{10}{\kHz}. %For systems with lower sampling rate, the total time of simulation should be longer. This makes sure the filter length remain in same order, such that the accuracy of filter is promised.  

The requirements of the closed-loop controlled system are given as
\begin{itemize}
	\item Rising time, $t_r < $\SI{5}{ms}
	\item Settling time, $t_s <$\SI{10}{ms}
	\item Maximum overshoot, $M_p < 1\%$
	\item Phase margin, $PM > \ang{80}$
	\item Steady-state error, $e_{ss} < 2\%$
\end{itemize}

For simplicity in design, assume that no overshoot occurs, i.e. $M_p=0\%$, hence the open-loop damping ratio $\zeta$ is \num{0}. The placement of the dominant pole can be evaluated by the approximation formula $t_r = 1.8/\omega_n$, corresponding to $\omega_n$ is \SI{360}{\radian / \s}. The closed-loop crossover frequency $\omega_c$ is estimated with $\omega_c =\omega_n/(1+\zeta^2)$, since $\zeta$ is chosen as \num{0}, the lower bound of $\omega_c$ is \SI{360}{\radian / \s}. On the other hand, the \SI{2}{\percent} steady state error indicates that the position constant $K_p$ of $L_d(z)$ should be higher than \SI{33.8}{\decibel}.

With aforementioned design constraints about $\omega_c$, $K_p$, the desired loop gain $L_d(z)$ is chosen to be a second-order system composing of a low-frequency pole to reduce the steady-state error and a lead compensator to ensure the phase margin specification.

\begin{equation} \label{eq:L_d}
    L_d(z) = \frac{0.3(z-0.9)}{(z-0.999)(z-0.7)}
\end{equation}

The closed-loop poles locate at \num{0.9444} and \num{0.4546}.
The design requirements and the frequency response are presented in Fig. \ref{fig:P}.

%--------------------------------------------------------------------------------
\subsection{ILC-Based Loop-Shaping}

To start with the ILC-based loop-shaping method proposed in this paper, the ILCFF with convergence rate acceleration algorithm should be performed to obtain an FIR inverse filter $f(k)$ of the plant $P(z)$. Note that the length of $f(k)$ is chosen as \num{5000}, which is \SI{0.5}{\s} corresponding to \SI{10}{\kHz} sampling rate. Specifically, half of the filter length is for the causal part and the other half for non-causal. The algorithm is run for a total of \num{100} iterations with the learning filter updating every \num{10} iterations. The resulting RMS tracking error described in Eq. \ref{eq:min_time} is around \num{2e-13}, where $r(k)=\delta(k)$ is a delta impulse function.

With the FIR inverse filter $f(k)$, the proposed loop-shaping method is then applied for learning the control input $c(k)$ using Eq. \ref{eq:ilc_ls}. The length of $L_d$(k) is truncated to \num{5000}  (\SI{0.5}{\s} at \SI{10}{\kHz} sampling rate).

\begin{figure}[t]
    \begin{center}
    \includegraphics[width=8.4cm]{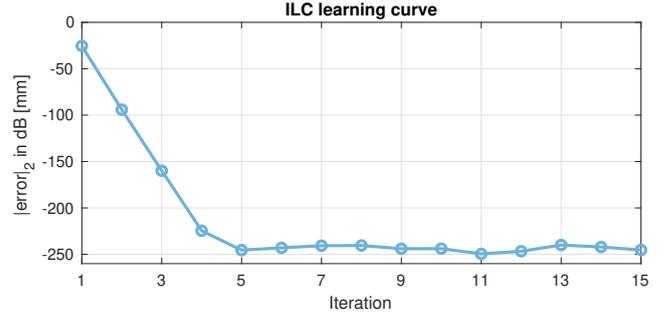}    % The printed column width is 8.4 cm.
    \caption{The ILC learning curve of tracking $l_d(k)$, in terms of the $l^2$-norm error $\abs{l_d(k)-c(k)P(z)}_2$.} 
    \label{fig:Learning Curve}
    \end{center}
\end{figure}

\begin{figure}[t]
    \begin{center}
    \includegraphics[width=8.4cm]{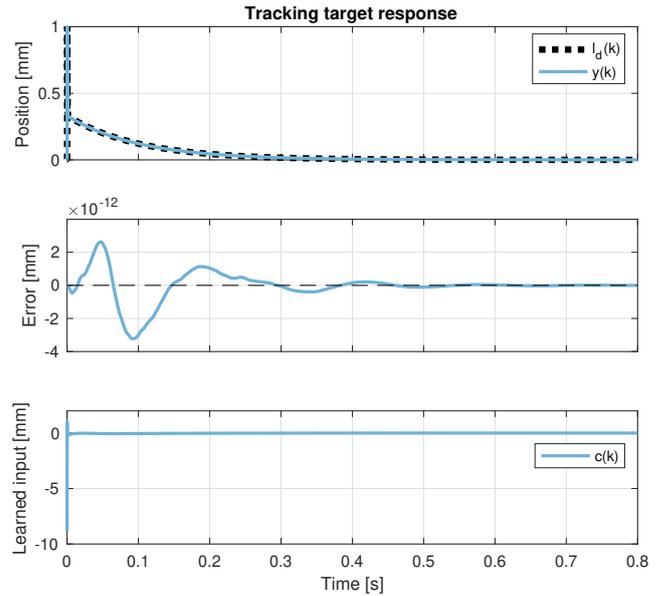}    % The printed column width is 8.4 cm.
    \caption{The converged iterative result of tracking $l_d(k)$ and the corresponding learned input $c(k)$.} 
    \label{fig:Tracking}
    \end{center}
\end{figure}

The simulation result of the $l^2$-norm tracking error convergence curve is shown in Fig. \ref{fig:Learning Curve}. The tracking error converges to around \SI{-240}{\decibel} at the \num{5}-th iteration, as the error is mostly contributed by the numerical error and hence the error stays around the same level after running \num{5} iterations. The iterative learning result $l_{\text{FIR}}(k)$ of tracking $L_d(k)$ and the corresponding leaned controller input $c(k)$ are presented in Fig. \ref{fig:Tracking}. The maximum tracking error is around \num{3e-12}, which is also a numerical error.

\begin{figure}[t]
    \begin{center}
    \includegraphics[width=8.4cm]{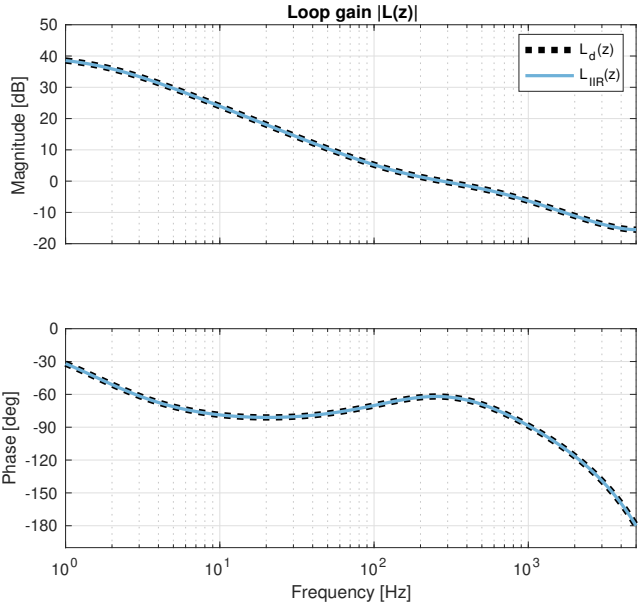}    % The printed column width is 8.4 cm.
    \caption{The frequency response of the desired loop gain and the loop gain learned with ILC.} 
    \label{fig:L}
    \end{center}
\end{figure}

\begin{figure}[t]
    \begin{center}
    \includegraphics[width=8.4cm]{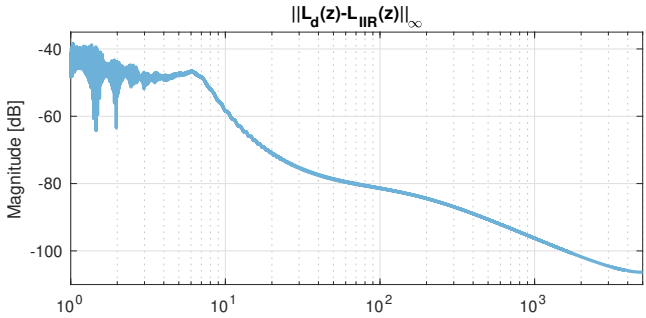}    % The printed column width is 8.4 cm.
    \caption{The frequency response of $\norm{L_d(z)-L_{\text{IIR}}(z)}_\infty$.} 
    \label{fig:reduction}
    \end{center}
\end{figure}

To perform a real-time feedback control, the learned filter $c(k)$ is converted to an IIR filter $C_{IIR}(z)$ through balanced realization. By investigating the Hankel singular values of the FIR filter, the model can be reduced to a \num{5}-th order controller, which consists of \SI{96}{\percent} of the summation of original Hankel singular values. The IIR-type loop gain $L_{IIR}(z)$ is formed by Eq. \ref{eq:L_IIR}. The error bound of $\norm{C_{\text{FIR}}(z)-C_{\text{IIR}}(z)}_\infty$ can be obtained from Eq. \ref{eq:err_bound}, where $r$, $m$ are the reduced model order \num{5} and the original model order \num{10000}, respectively. The error bound is promised by Hankel singular value according to Eq. \ref{eq:err_bound}, which is 5.52 or \SI{14.84}{\decibel}, as the maximum error is around only \SI{-38}{\decibel}.
Finally, Fig. \ref{fig:L} plots the desired loop gain and the resulting loop gain $L_\text{IIR}(z)$,indicating that the learned loop gain fits the desired loop gain both in magnitude and phase. The error between the two loop gains is less than \SI{-38}{\decibel} in all frequency presented in Fig. \ref{fig:reduction}.

%--------------------------------------------------------------------------------
\subsection{Closed-Loop Validation}

The closed-loop step response is simulated to identify whether the controller meets the design specification. Eq. \ref{eq:close-loop} is used to obtain the closed-loop system for the desired loop gain, learned FIR and IIR loop gain, which are $G_d(z)$ and $G_{\text{IIR}}(z)$, receptively.

\begin{equation} \label{eq:close-loop}
    G(z) = (I+C(z)P(z))^{-1}C(z)P(z)
\end{equation}

\begin{figure}[t]
    \begin{center}
    \includegraphics[width=8.4cm]{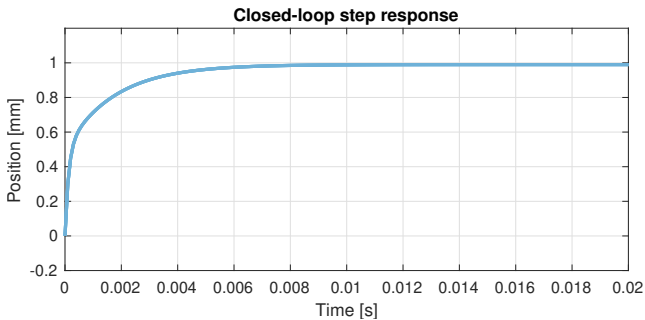}    % The printed column width is 8.4 cm.
    \caption{The step response of the closed-loop system.} 
    \label{fig:CL}
    \end{center}
\end{figure}

The simulation results of the closed-loop step response are shown in Fig. \ref{fig:CL}. Specific values of the step response properties are list in Tb. \ref{tb:spec}. Each specification of $G_d(z)$ and $G_{\text{IIR}}(z)$ are equal up to \num{4} decimal digits. Therefore the controller learned from the method proposed in this paper meets the provided requirements.

\begin{table}[t]
    \begin{center}
    \begin{tabular}{ccccc}
        \hline
        & spec. & $G_d$ & $G_{IIR}$\\\hline
        $t_r(ms)$ & 5 & 2.7 & 2.7\\
        $t_s(ms)$ & 10 & 5.7 & 5.7\\
        $M_p(\%)$ & 1 & 0 & 0 \\
        $PM(\si{\deg})$ & 80 & 118 & 118 \\
        $e_{ss}(\%)$ & 2 & 0.99 & 0.99
        \\ \hline
        \\
    \end{tabular}
    \caption{The step response simulation results and validation of requirements.}\label{tb:spec}
    \end{center}
\end{table}

%================================================================================
\section{Conclusion}
\label{sec:5}

A novel method for model-free loop-shaping based on ILC is proposed in this paper. The conventional loop-shaping problem is converted into a tracking problem, which is solved by the ILC algorithm, taking advantage of its tracking performance and model-free process. While the learning for the impulse response of the desired loop gain is done, the learned control input is used to construct an FIR filter as the closed-loop controller. For real-time feedback implementation, the FIR controller is then further converted into an IIR filter by balanced realization. The proposed is validated on a 3-rd order numerical plant. The simulation results show that the constructed controller meets the desired loop gain, as the learning process converges within only a few iterations.

%================================================================================
\bibliography{ifacconf} 

\end{document}